\documentclass{vgtc}                         
\ifpdf
  \pdfoutput=1\relax                   
  \usepackage{graphicx}                
  \DeclareGraphicsExtensions{.pdf,.png,.jpg,.jpeg} 
\else
  \usepackage{graphicx}                
  \DeclareGraphicsExtensions{.eps}     
\fi%

\graphicspath{{figures/}{pictures/}{images/}{./}} 

\usepackage{microtype}                 
\PassOptionsToPackage{warn}{textcomp}  
\usepackage{textcomp}                  
\usepackage{mathptmx}                  
\usepackage{times}                     
\usepackage{cite}                      
\usepackage{tabu}                      
\usepackage{booktabs}                  

\usepackage{xcolor}

\onlineid{0}

\vgtccategory{Research}

\title{Using Abstract Tangible Proxy Objects for Interaction in Optical See-through Augmented Reality\vspace{20pt}}

\author{Denise Kahl\thanks{e-mail: denise.kahl@dfki.de} %
\and Antonio Krüger\thanks{e-mail: antonio.krueger@dfki.de}}
\affiliation{\scriptsize German Research Center for Artificial Intelligence  (DFKI), \\ Saarland Informatics Campus, Saarbrücken, Germany}

\abstract{
Interaction with virtual objects displayed in Optical See-through Augmented Reality is still mostly done with controllers or hand gestures. A much more intuitive way of interacting with virtual content is to use physical proxy objects to interact with the virtual objects. Here, the virtual model is superimposed on a physical object, which can then be touched and moved to interact with the virtual object.
Since it is not possible to use an exact replica as a tangible proxy object for every use case, we conducted a study to determine the extent to which the shape of the physical object can deviate from the shape of the virtual object without massively impacting performance and usability, as well as the sense of presence.
Our study, in which we investigated different levels of abstraction for a sofa model, shows that the physical proxy object can be abstracted to a certain degree. At the same time, our results indicate that the physical object must have at least a similar shape as the virtual object in order to serve as a suitable proxy.
} 

\begin{document}



\keywords{Tangible augmented reality, optical see-through augmented reality, tangible interaction, haptic devices.}

\CCScatlist{
   \CCScatTwelve{Human-centered computing}{Human computer interaction (HCI)}{Interaction paradigms}{Mixed / augmented reality};
   \CCScatTwelve{Human-centered computing}{Human computer interaction (HCI)}{Interaction devices}{Haptic devices}{}
 }

\teaser{
    \includegraphics[width=\textwidth]{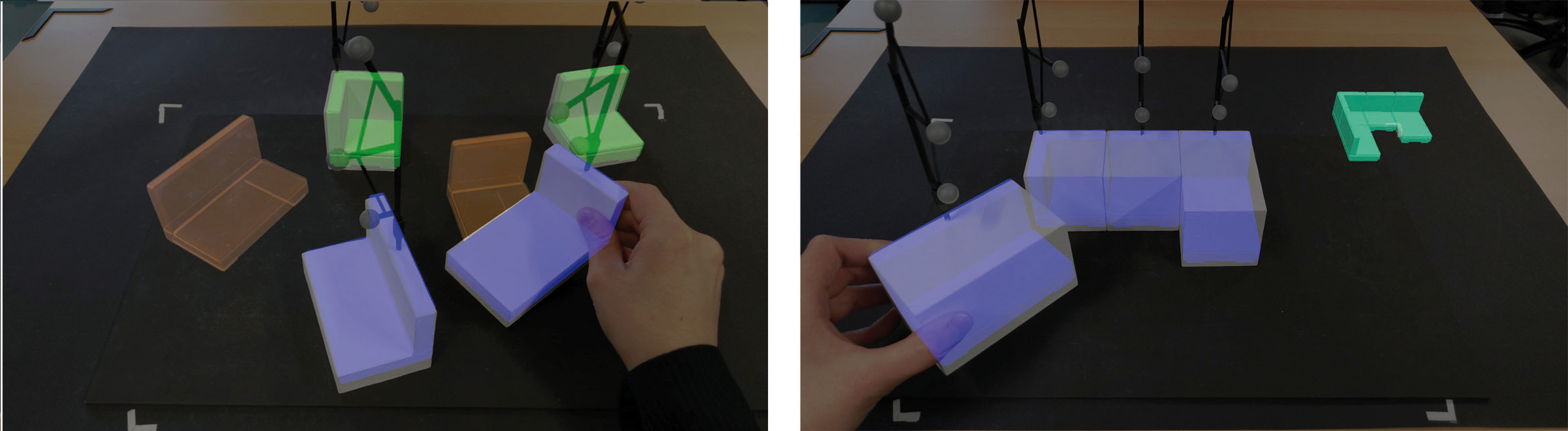}
    \caption{Left: HoloLens screenshot of study part 1 in condition~\textit{\textbf{B}}: Matching virtual sofa parts (blue) to appropriate 3D targets (orange). Correctly matched objects are displayed in green. Right: HoloLens screenshot of study part 2 in condition~\textit{\textbf{D}}: Assembly of the virtual sofa parts (blue) according to a 3D miniature template (turquoise).}
    \label{fig:teaser}
}

\maketitle

\section{Introduction}
Augmented Reality (AR) technology serves as a bridge between the real world and the virtual realm by overlaying virtual objects onto the user's field of view (FOV)\cite{billinghurst2008tangible,VanKrevelen2010,Azuma1997}. Its applications span across diverse fields, including but not limited to medicine, where it has found utility in surgical applications\cite{vavra2017recent}; education, where it has been employed to enhance learning experiences~\cite{chen2017review}; and architectural and urban design, where it has facilitated the visualization and planning of structures~\cite{sorensen2013development}.

The evolution of AR technology has led to the emergence of Optical See-through AR (OST AR) applications, which offer users an almost unobstructed view of the real world, as opposed to Video See-through AR (VST AR). This has opened up new possibilities for interaction and immersion in augmented environments~\cite{Rolland2000}. However, it is important to note that the transparency and opacity of the overlays in OST AR are influenced by the surrounding lighting conditions~\cite{erickson2020exploring, kahl2022influence}, necessitating a distinct examination of this modality separate from VST AR.

One significant advantage of using AR headsets is the ability to interact with physical objects using two free hands. When these objects are used as physical proxies and coupled with digital information~\cite{Ishii1997}, manipulation tasks involving virtual objects can be performed more efficiently and with greater precision~\cite{besanccon2017mouse, tuddenham2010graspables}. This concept, known as Tangible Augmented Reality (TAR), enables intuitive and natural interaction with virtual content and has led to the exploration of various applications and use cases~\cite{billinghurst2008tangible, choi2019applying}. The benefits of TAR can also be harnessed in the context of OST AR headsets.

While many AR applications primarily rely on controllers, hand gestures, or speech for interaction~\cite{munsinger2019usability, chang2017evaluating, liu2018technical}, the incorporation of tangible physical objects into the AR experience introduces novel opportunities. For example, in the field of architecture, users can go beyond mere 3D object visualization and actively collaborate to create or modify virtual buildings.
In the world of gaming, the incorporation of physical objects, such as interactive game characters, enriches the overall gaming experience by providing a higher level of interactivity and immersion through tactile interaction~\cite{hoffman1998physically}.

To enable intuitive and versatile interaction in a wide range of applications, a vast array of physical props would be required. However, creating and storing exact physical replicas for every virtual object in every possible use case is impractical~\cite{de2019different, Hettiarachchi2016, DeTinquy2019}. As a more feasible approach, the use of abstract physical objects that can represent multiple virtual objects becomes essential. Ideally, users would have access to a collection of abstract props that can be adapted for various scenarios. To determine the acceptable range of differences in size, shape, texture, and material between the physical props used for interaction and their corresponding virtual objects, thorough investigations are necessary. These studies will provide valuable insights into achieving a seamless integration of physical and virtual elements in AR experiences.

There are already several investigations in OST AR that have examined the extent to which a physical object used to interact with virtual content can differ in size from its virtual counterpart~\cite{kahl2021investigation, kahl2022influence}. To our knowledge, no studies have yet been conducted on feasible shape deviations between physical and virtual objects in OST AR.

In this paper, we use a variety of abstraction levels to investigate how much the shape of the physical proxy object can differ from the shape of the virtual object in OST AR without leading to significantly worse performance and usability ratings, and without creating a significantly worse sense of presence. In our study, we have used sofa models as an example, whose physical models have been abstracted into five different variants.

\section{Related Work}
There is already some research on the extent to which a virtual object and a physical proxy object can differ. Many of these investigations have been carried out in VR.

Simeone et al.~\cite{simeone2015substitutional} conducted a study to determine the acceptable level of discrepancy between physical and virtual objects in virtual reality (VR) without breaking the VR illusion. They had participants interact with different virtual objects in a VR environment that mimicked a medieval courtyard. Each virtual object had a corresponding physical counterpart in the real world that participants could touch and manipulate.
The study examined various substitutions of a mug in VR, including a matching virtual replica, a virtual model with aesthetic differences, a model with added or omitted parts, a functionally different model, and a virtual model with categorical differences. The researchers found that an exact replica of the mug in VR was significantly more believable than versions with shape or visualized temperature differences. They also noted that virtual objects smaller than their physical counterpart had a lower level of credibility compared to larger virtual representations.

In another study, de Tinguy et al.~\cite{de2019different} investigated the similarity required between virtual and physical objects for them to feel the same. Participants were asked to touch a certain object while the virtual overlay changed based on variations in width, local orientation, and local curvature. The study found that participants could perceive differences in local curvature at a discrepancy of 66.66\%, while orientation differences were not detected up to a discrepancy of 43.8\%. The study also revealed that the width of objects could be changed by up to 5.75\% in VR without users noticing a difference.

Bergström et al.~\cite{bergstrom2019resized} discovered that smaller and larger physical objects could be used as proxies for interacting with virtual objects without users noticing the size difference in VR. They developed a method called "Resized Grasping" that manipulated virtual finger positions to achieve this effect. However, such methods are applicable only in purely virtual environments and cannot be applied to AR, where users see their real hands.

Considering the findings from these studies, Nilsson et al.~\cite{nilsson2021haptic} established three criteria for successfully using proxy objects in VR: sufficient similarity, complete co-location, and compelling contact forces. Proxy objects should closely match the haptic properties of the virtual objects in terms of shape, size, and weight. It is important for the physical objects to be co-located with their virtual counterparts, and appropriate stimuli should be provided when forces are applied to objects in the virtual environment. The importance of these criteria varies depending on the specific VR application.

Also in VST AR, there are already initial investigations regarding feasible differences between physical proxy objects used for interaction and their virtual representations.

Kwon et al.~\cite{kwon2009effects} carried out the first studies on size and shape variations in VST AR. They conducted a study using three tangible objects with significant differences in shape and size. The study revealed that participants performed best when the shape and size of the virtual object matched the corresponding physical object. The largest time differences were observed during the grasping phase, while the subsequent manipulation time did not show significant variations. In an additional experiment that focused solely on size variations, no significant performance differences were found among the five different size conditions. Based on these findings, the researchers concluded that the shape differences were primarily responsible for the results observed in the main study.

Kobeisse and Holmquist~\cite{kobeisse2022can} explored the design possibilities of using generic objects as substitutes for historical artifacts within the context of cultural heritage in VST AR. In their user study, participants interacted with a virtual 3D model of a Bronze Age urn using four different interfaces: a touch screen, a flat AR marker, a generic wooden cylinder, and a 3D-printed replica of the digital artifact. The findings show that the 3D-printed replica, which closely resembles the physical object, offers the most realistic method of interaction. However, the study also suggests that using a generic object like the wooden cylinder can provide a more immersive experience when compared to the touch screen and flat AR marker. Overall the results indicate that tangible interfaces enhance engagement and offer a more authentic experience compared to traditional observation without tactile interaction.

In addition to the described investigations in VR and VST AR, there are also already first investigations on the use of everyday proxy objects as well as on possible size differences between the virtual and the physical object in OST AR.

Hettiarachchi and Wigdor~\cite{Hettiarachchi2016} introduced an approach that explores the concept of using everyday objects as tangible interfaces in OST AR. Their system scans the environment to identify real-world objects that closely resemble virtual objects and then overlays virtual models onto these identified objects in AR. 
Similarly, Szemenyei and Vajda~\cite{szemenyei2015learning, szemenyei20183d} developed algorithms that enable automatic matching of everyday physical objects with virtual objects.

Kahl et al. investigated the effects of size variations as well as the influence of ambient lighting on size variations in OST AR~\cite{kahl2021investigation, kahl2022influence}. They conducted experiments with a tangible AR setup in which users could interact with virtual objects superimposed on physical objects. Participants were presented with virtual objects of various sizes, and participants' responses and performance on various tasks were analyzed. The studies found that the physical and virtual objects can differ in size to some degree without degrading usability and performance. In this context, for physical objects that are smaller than the virtual object, stronger size deviations are possible than for larger physical objects. The size ranges in which a physical object can deviate from its virtual counterpart increase with increasing ambient illumination due to the reduced opacity of virtual overlays and thus better visibility of the physical proxy.

To our knowledge, there are no previous studies in the field of OST AR that examine shape differences between physical proxy objects and their corresponding virtual counterparts.
Therefore, the purpose of this paper is to explore the impact of shape variations of proxy objects on the perception of reality, ease of use, and performance in OST AR. Our aim is to determine the feasibility of using abstracted proxy objects for interacting with virtual content by examining various levels of abstraction.

\begin{figure}
  \centering
  \includegraphics[width=\linewidth]{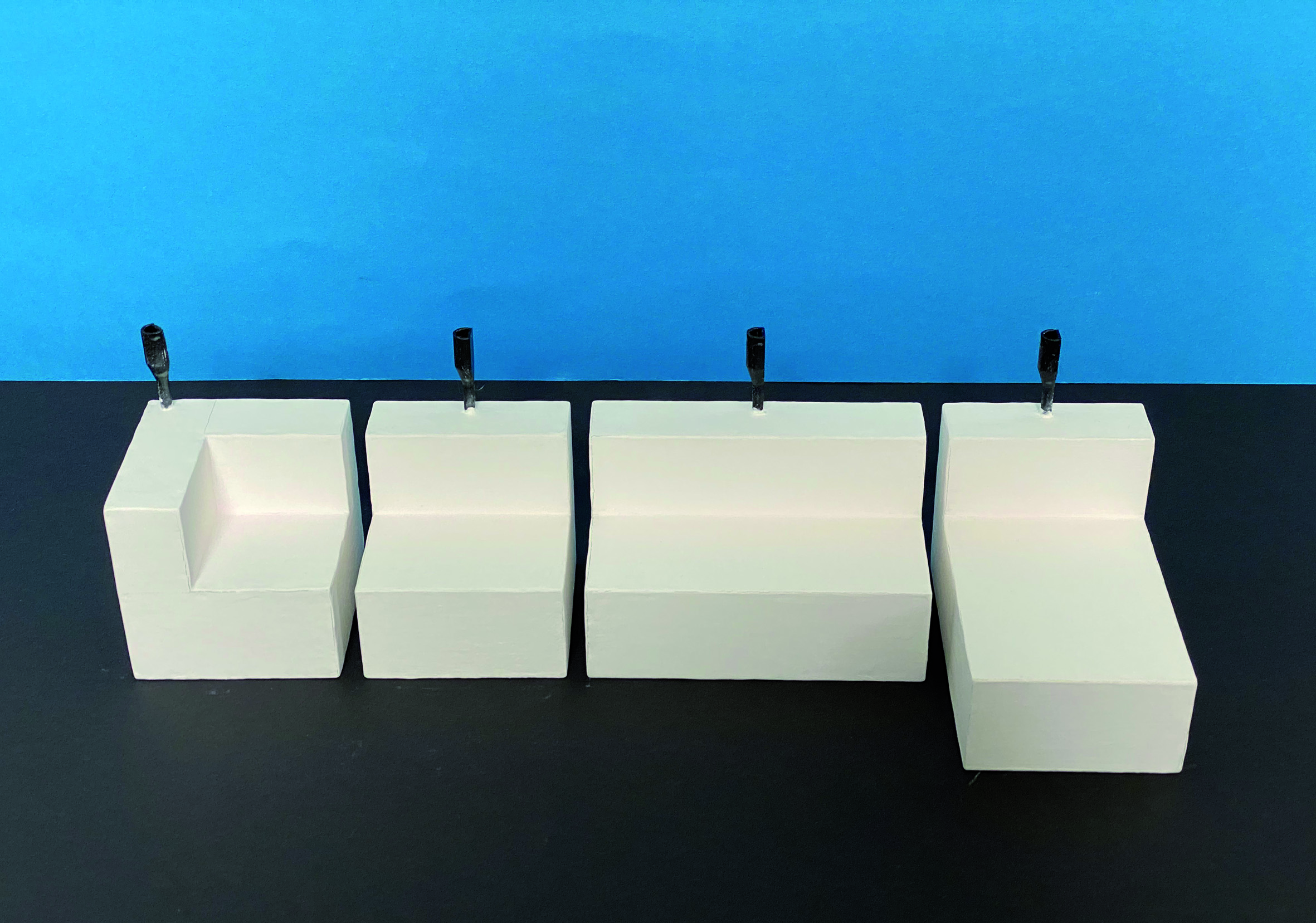}
  \caption{Sofa parts used in the study. From left to right: corner piece, one-seater, two-seater and chaise longue in shape condition \textit{\textbf{C}}.}
  \label{fig:sofa_parts}
\end{figure}

\section{Study}
A user study was conducted to find out whether it is possible to use abstracted proxy objects for interacting with virtual objects. Since the studies on size differences in OST AR~\cite{kahl2021investigation, kahl2022influence} have shown that small size differences are hardly perceived and therefore have little impact on usability (disturbance ratings), presence (AR/TAR presence), and performance (task completion time), we assumed that this would also be the case for very small shape variations. However we thought, that there would be strong negative effects above a certain degree of difference.
Similarly, we assumed that there could be a strong deterioration in performance if there were very large differences in shape.
Therefore, the following hypotheses were made:

\begin{itemize}
  \item[H1:] Shape differences between virtual and physical objects cannot be properly detected for very similar shapes.
  \item[H2:] The shape of the virtual and physical object can differ within a certain range without significantly degrading ``AR presence'' and ``TAR presence''.
  \item[H3:] The shape of the virtual and physical object can differ within a certain range without significantly worsening usability.
  \item[H4:] The level of difference between the shape of the virtual and physical object has an impact on performance.
\end{itemize}

\subsection{Tasks}
To test the above hypotheses, we defined two study parts in which each virtual object was represented by five different physical props that served to interact with it. In our study, we used four different sofa parts as interaction objects~(see Figure~\ref{fig:sofa_parts}).

In each part of the study, the tasks had to be solved as quickly and as precisely as possible.
One part of the study consisted of a 2D alignment task. Here, the virtual 3D objects had to be independently adjusted to the respective matching 3D target objects, which were displayed at an appropriate distance as an AR overlay on the workspace (see Figure~\ref{fig:teaser}, left).
The other part of the study was to assemble the individual virtual objects into a composite object anywhere on the workspace based on a given miniature 3D model displayed in the upper right corner of the AR view (see Figure~\ref{fig:teaser},~right).

In both parts of the study, we made sure to include interactions with multiple objects so that the props would need to be grasped and moved more often, for example, to better measure the influence of the shape of the physical object when grasping~\cite{kwon2009effects}. All tasks consisted of simple subtasks, such as grasping, lifting, moving, and placing, all of which are necessary to solve high-level tasks~\cite{MacKenzie99}.

\subsection{Participants}
20 study participants were recruited (4 female, 16 male) in an age range from 19 to 51 years ($M=26.95, SD=8.204$). Those outside our institution received 10 Euro for their participation.
19 participants were right-handed, one was left-handed, and all of them had normal or corrected-to-normal vision.
On a 7-point Likert scale, participants reported relatively low experience with AR ($M=2.35, SD=1.309$) and very low experience with AR glasses ($M=1.25, SD=0.444$) where 1 means they had never used such systems and 7 means they used them regularly.

\subsection{Setup}
The study took place in a darkened room to ensure that lighting conditions during the study were not influenced by external factors (such as sunlight) and that the same lighting conditions prevailed for all participants.
The room was illuminated with the fluorescent tubes of the ceiling lighting. The measured luminance on the tabletop facing upwards was $98.5lx$ and from the camera of the Head Mounted Display (HMD) looking diagonally downwards onto the interaction surface $14.8lx$. This medium brightness was chosen as it is quite natural but not too bright to still be able to see the overlays well.

\begin{figure}
  \centering
  \includegraphics[width=\linewidth]{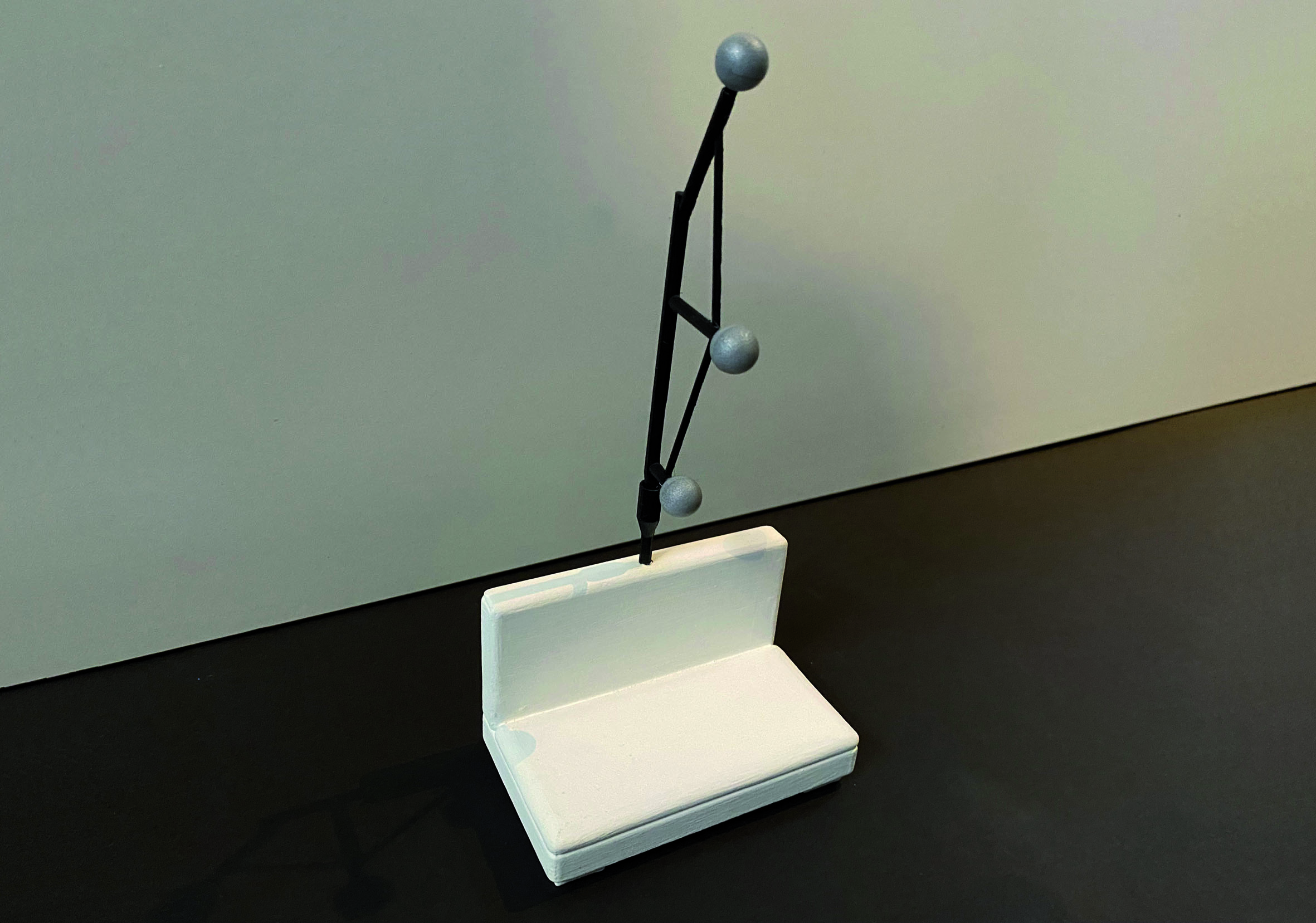}
  \caption{Two-seater in shape condition \textit{\textbf{A}} with attached marker tree.}
  \label{fig:twoseater_markertree}
\end{figure}

For tracking of the physical proxy objects we used 10 OptiTrack Prime\textsuperscript{X}~13 cameras\cite{Optitrack}. These were installed on a truss with a dimension of about 4~m~x~6~m and directed to the center of the tracking area. The table where the participants sat while working on the tasks was located in the center of this area.
We equipped the physical props as well as the HMD with marker trees for tracking (see Figure~\ref{fig:twoseater_markertree}).
The virtual objects were displayed as AR overlays using a HoloLens~2~\cite{Microsoft}, with the corresponding application implemented in Unity~\cite{Unity} (version 2019.3.8f1).

The tasks were performed on a black wooden board on a black background, which was large enough so that both the physical and the virtual objects were always visible against this black background.

The objects interacted with were white with black marker trees for tracking (see Figure~\ref{fig:twoseater_markertree}).
The HoloLens~2 used for the study was set to maximum brightness and the opacity of the overlays was set to 100\%. For the color of the overlays we chose a medium-dark blue (\#001FFF), since a pre-test showed that under the given lighting conditions, the virtual and physical object are visible equally well with this color choice.
Participants were placed at the table so that they all had approximately the same viewing angle (approx.\ 45 degrees) of the objects they had to interact with to solve the tasks.

\begin{figure*}
  \centering
  \includegraphics[width=\linewidth]{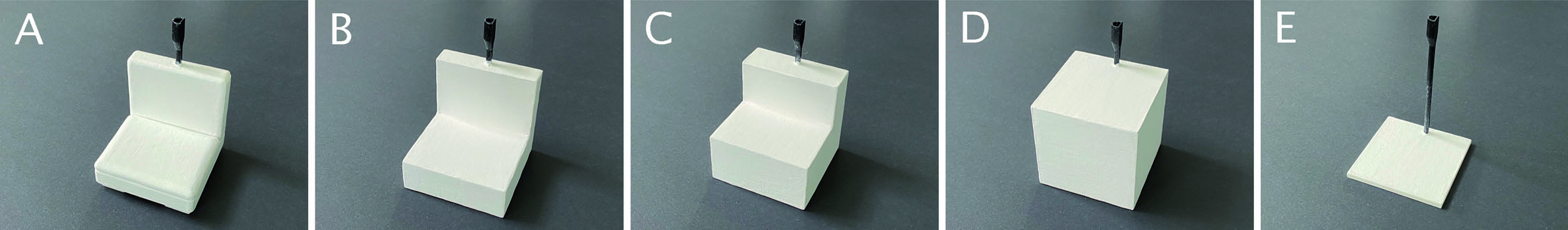}
  \caption{Investigated shape variations of the physical proxy object. Shape conditions left to right: 3D model (\textit{\textbf{A}}), abstracted 3D model (\textit{\textbf{B}}), abstraction of a standard sofa (\textit{\textbf{C}}), cuboid (\textit{\textbf{D}}) and plane (\textit{\textbf{E}}).}
  \label{fig:shapes_shape}
\end{figure*}

\begin{table*}
  \centering
  \caption{Seat heights and backrest depths of the different shape variations in centimeters.}
  \label{tab:sofa_measures}
  \begin{tabular}{l c c c c c}
    \toprule
    &   & Abstracted & Abstraction of &   &   \\
    & 3D Model & 3D Model & Standard Sofa & Cuboid & Plane\\
    & \textit{\textbf{A}} & \textit{\textbf{B}} & \textit{\textbf{C}} & \textit{\textbf{D}} & \textit{\textbf{E}}\\
    \midrule
    Seat Height & 2.1 & 2.1 & 3.2 & 6.0 & 0.3 \\
    Backrest Depth & 1.2 & 1.2 & 2.0 & 6.0 & 0.0 \\
  \bottomrule
\end{tabular}
\end{table*}

Whenever a task was solved the participants had to complete three different questionnaires: an AR presence questionnaire, a TAR presence questionnaire, and a shape perception questionnaire.
The AR presence and TAR presence questionnaires were used to assess how realistic the overlays appeared in each task and how realistic the interaction with the virtual objects felt, respectively.
To ensure comparability of results with related work, we used the same AR presence and TAR presence questionnaires as Kahl et al. used in their study on the influence of lighting conditions on size variations and had them answered in AR using a tangible proxy pen~\cite{kahl2022influence}.
The shape perception questionnaire consisted of three questions. It assessed the perceived shape differences between the virtual and the physical object on a scale from 1 (= not fitting at all) to 7 (= completely fitting) as well as the disturbance during grasping and interaction. All questionnaires were rated using 7-point Likert scales.

After finishing all tasks, the participants received a final paper questionnaire asking for demographic information and additionally for a ranking of the different physical shapes displayed in Figure~\ref{fig:shapes_shape}.

\subsection{Basic Approach}
At the start of each study session, the HoloLens~2 was adjusted to the eyes of the corresponding participant. This step was necessary because people have different depth perception due to factors such as interpupillary distance. The adaptation process involved two steps. First, the participants performed the eye calibration of the HoloLens~2. Then, this calibration was checked using a sample object. In cases where the virtual object overlaid on the physical proxy object did not align accurately, the position of the virtual object was manually adjusted in 3D using our Unity application until a perfect match was achieved.

This adjustment was especially necessary for participants who had thick glasses or a very wide interpupillary distance. Performing the eye calibration ensured precise alignment of the overlays regardless of the position of the physical object during the study.

A systematic procedure was followed for each task. Initially, participants were prompted to align the FOV of their HoloLens by matching a designated area on the table with a white frame displayed in the AR interface. Participants maintained this position to guarantee continuous visibility of the overlays on the physical objects within their FOV.

Throughout this process, the physical props were carefully arranged on a plate concealed from the participants' view. The plate was covered with a box and then positioned at a specific location on the table in front of the participant. Only after removing the cover box symbolizing the beginning of the task, the overlays on the objects and the virtual targets or the virtual template became visible. Simultaneously, the automatic measurement of time was initiated.

\subsection{Design and Procedure}
The study was designed as a within-subjects experiment. In total, five different shape conditions were tested.
The order of the shape conditions in both parts of the study was counterbalanced by a Williams design Latin square of size five~\cite{williams1949experimental}. Half of the participants started with the first part of the study and the other half with the second part of the study.
Figure~\ref{fig:shapes_shape} shows the shape variations of the physical proxy objects. Condition \textit{\textbf{A}} represents the baseline where the physical object is an exact replica (3D model) of the virtual object. Condition \textit{\textbf{B}} is an abstraction of the 3D model, where seat height and seat depth match those of the virtual object. Condition \textit{\textbf{C}} is an abstraction of a standard sofa. To determine the standard dimensions, the seat heights and seat depths of all available IKEA~\cite{IKEA}
sofas were averaged. Condition \textit{\textbf{D}} (Cuboid) corresponds to the bounding box of the sofa part and condition \textit{\textbf{E}} to the base area. Except for condition \textit{\textbf{E}} -- because it is flat -- the external dimensions correspond to those of the virtual object, so that only the varied shapes have an influence on the results.
The seat heights and backrest depths of the different abstractions can be found in Table~\ref{tab:sofa_measures}.

\begin{figure*}
  \centering
  \includegraphics[width=\linewidth]{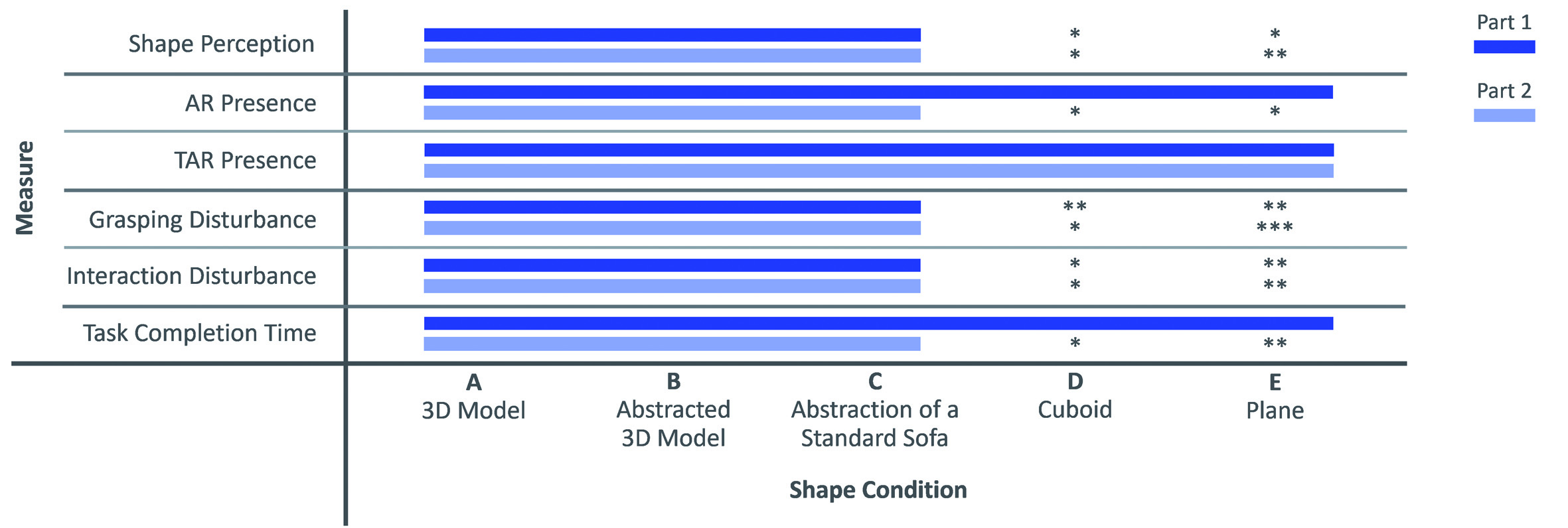}
  \caption{Summary of significant differences of the shape conditions compared to the shape-matching condition \textit{\textbf{A}} as a baseline marked with *~($p<0.05$), **~($p<0.01$) and ***~($p<0.001$) for the two study parts. The blue bars indicate the ranges without significant difference.}
  \label{fig:shape_summary}
\end{figure*}

The proxy objects for abstractions \textit{\textbf{B}}, \textit{\textbf{C}}, \textit{\textbf{D}} and \textit{\textbf{E}} were laser-cut from MDF boards for environmental reasons. They were assembled into the sofa pieces and then painted several times with white paint. The 3D models of abstraction \textit{\textbf{A}} were 3D printed and afterwards painted several times with the same white paint to guarantee the same feeling when touching the objects.
In addition to the surface, care was also taken to ensure that the individual sofa parts had almost the same weight (+/- 1 gram) in all abstraction conditions (except condition \textit{\textbf{E}}), as well as a similar weight distribution (main weight in the lower back area; except conditions \textit{\textbf{D}} and \textit{\textbf{E}}).
All sofa parts were equipped with adapters on which the appropriate marker trees could be mounted. In this way, only four different marker trees with three reflective markers each were needed. For each physical sofa part, a fine calibration of the corresponding virtual object was performed in Unity to ensure an accurate overlay of the virtual object.
A total of 20 different sofa parts was produced in this way, with which the participants interacted during the study.

In part 1 of the study, the sofa parts had to be arranged individually. Virtual 3D targets to which the virtual overlays of the physical sofa parts had to be matched were displayed at four predefined positions.
The physical objects were randomly arranged at the bottom of the plate. Likewise, the mapping of the virtual sofa parts to the four positions as well as the rotation of the virtual target objects was randomized.
To ensure that the rotation of the targets was always clearly visible without the need to move one's head, care was taken to ensure that the sofa parts were never shown predominantly from behind. Therefore, we excluded a direct view of the back as well as rotations up to 60 degrees in any direction from it.
In all five shape conditions, the four different objects had to be interacted with. The task was to align all four virtual overlays with the displayed 3D targets (see Figure~\ref{fig:teaser}, left). The order in which the objects could be assigned was flexible. As soon as an object was aligned precisely enough, its overlay turned green. The task was completed as soon as all overlays were green.
An object was considered correctly aligned if errors below a threshold of 4~mm in horizontal distance, 5~mm in vertical distance and 3° in the angle between virtual overlay and virtual target were detected constantly for 0.5 seconds.
Pre-tests showed that the choice of these values made the task sufficiently complex, but still feasible without too much effort.

In the second part of the study, the four sofa pieces had to be put together to form a specific sofa combination.
Also in this task, the physical objects were randomly arranged at the bottom of the plate at the beginning of each trial and randomly rotated.
The sofa combination that had to be recreated was displayed as a virtual 3D object in small format in the upper right area on the plate (see Figure~\ref{fig:teaser}, right).
It could be built anywhere on the plate and rotated as desired.
Once the sofa combination was correctly assembled, all overlays turned green at the same time and the task was completed.
Two individual elements were considered to be correctly assembled if they were connected with the correct sides and these had a maximum distance of 2~mm from each other and were at an angle of less than 7.5° to each other. In addition, the offset of the contact edges had to be less than 6~mm when aligned parallel to each other. As soon as these conditions were no longer met for 0.1 seconds, the connection between the parts was released.
We also determined these values in pre-tests to ensure that the task would not be too easy to solve, but also would not cause frustration among the participants.

\section{Results}
We investigated the effects of shape variations between the tangible and the corresponding virtual object on each of the measured dependent variables reported below with a uniform procedure. First, we applied a Friedman test with four degrees of freedom and a significance level of $\alpha=0.05$ to compare the samples collected in the five shape conditions with each other, separately for part 1 and part 2 of the study. We report the $p$-value along with the test statistic $\chi^2$. If a significant influence of the shape condition on the dependent variable was detected, we used Wilcoxon’s signed-rank tests ($dof=19$, $\alpha=0.05$) to compare each shape abstraction to condition \textit{\textbf{A}} (3D model) as a baseline. We report Bonferroni-corrected $p$-values, the test statistic $W$ and the matched pairs rank-biserial correlation $r$ as an effect size.
Figure~\ref{fig:shape_summary} shows an overview of our results obtained with the post-hoc tests.
Except for task completion time, no significant difference was found between part 1 and part 2 of the study when comparing the measures.

\subsection{Shape Perception}
The rating of to what extent the shapes of the virtual and physical objects matched was influenced significantly by which shape abstraction was used for both parts of the study. The Friedman test confirmed this for part 1 ($\chi^2=18.006, p=0.001$) and 2 ($\chi^2=34.183, p<0.001$). Wilcoxon’s signed-rank test revealed that in part 1 of the study, conditions \textit{\textbf{C}} (abstraction of a standard sofa) ($W=48, p=0.399, r=-0.439$) and \textit{\textbf{B}} (abstracted 3D model) ($W=46.5, p=1.0, r=-0.114$) did not differ significantly from the shape-matching condition \textit{\textbf{A}}, while abstractions \textit{\textbf{E}} (plane) ($W=20, p=0.018, r=-0.766$) and \textit{\textbf{D}} (cuboid) ($W=24, p=0.029, r=-0.719$) were found to match significantly less well to the virtual shape. Equally, in part 2, conditions \textit{\textbf{C}} ($W=22, p=0.664, r=-0.436$) and \textit{\textbf{B}} ($W=21.5, p=1.0, r=-0.218$) did not show a significant difference to \textit{\textbf{A}}, while conditions \textit{\textbf{E}} ($W=6, p=0.002, r=-0.93$) and \textit{\textbf{D}} ($W=24, p=0.03, r=-0.719$) did.
The results show that in our study setup slight differences of the proxy object compared to the virtual object were often not noticed.

\begin{figure*}[ht]
  \centering
  \includegraphics[width=\linewidth]{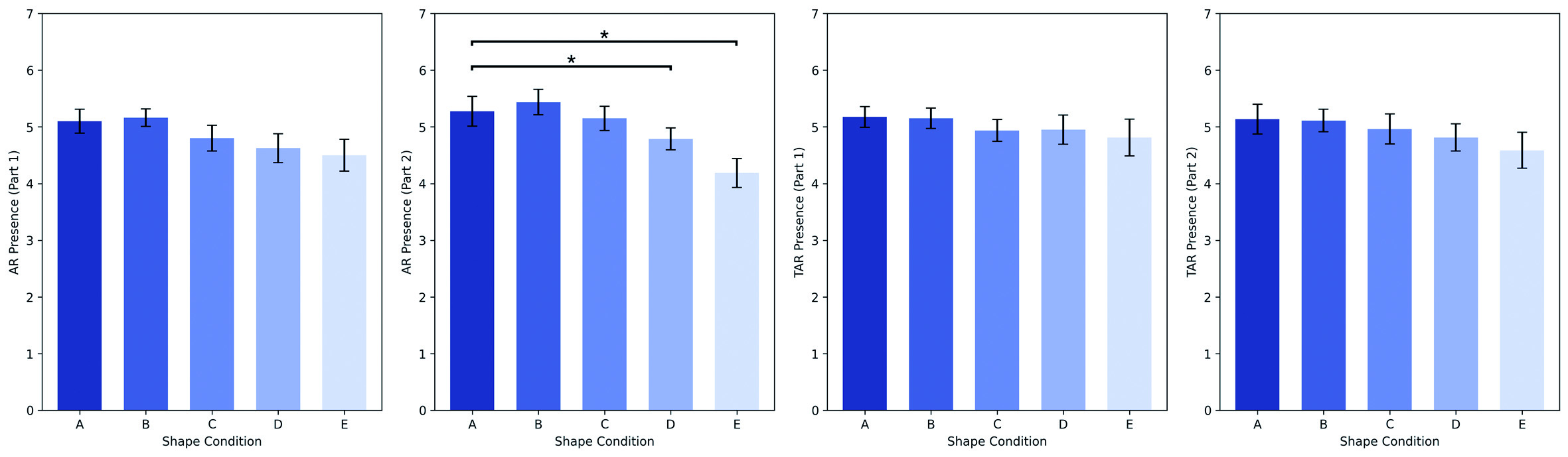}
  \caption{Mean AR presence and TAR presence ratings from 1 to 7 with marked standard errors for each shape condition. Significant differences from the baseline condition \textit{\textbf{A}} are marked with * ($p<0.05$).}
  \label{fig:shape_presence}
\end{figure*}

\begin{figure*}
  \centering
  \includegraphics[width=\linewidth]{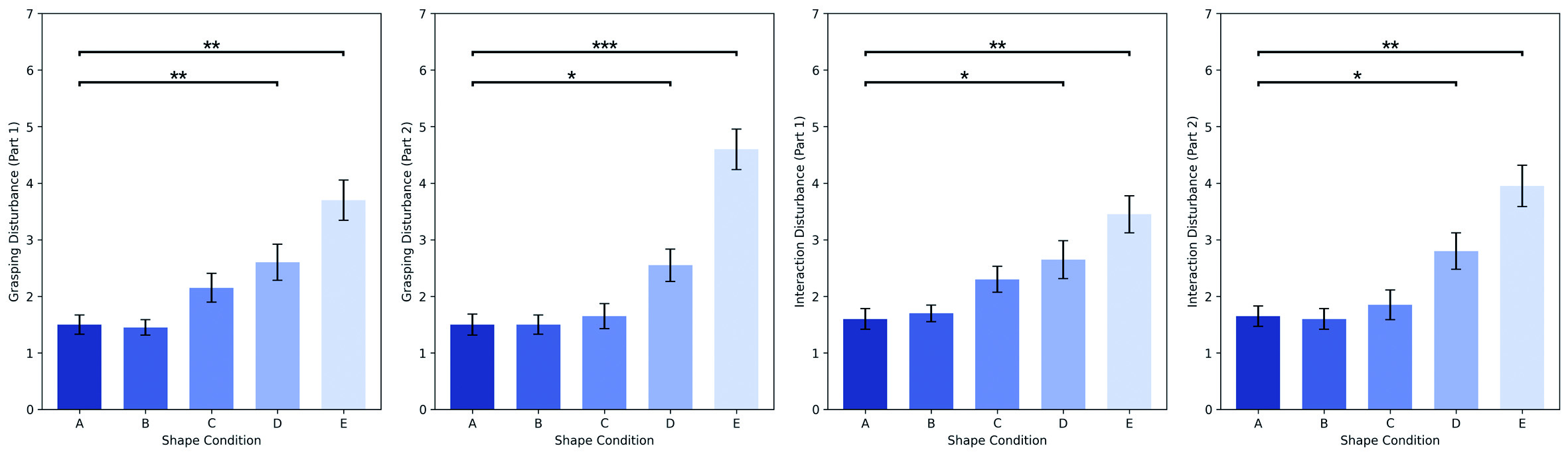}
  \caption{Mean disturbance ratings from 1 to 7 with marked standard errors for each shape condition. Significant differences from the baseline condition \textit{\textbf{A}} are marked with * ($p<0.05$), ** ($p<0.01$) and *** ($p<0.001$).}
  \label{fig:shape_disturbance}
\end{figure*}

\subsection{Presence}
According to the Friedman tests, AR presence scores are significantly influenced by the shape condition only in part 2 ($\chi^2=27.04, p<0.001$), not in part 1 ($\chi^2=3.024, p=0.554$). Wilcoxon’s signed-rank test revealed for part 2 that both conditions \textit{\textbf{E}} ($W=19, p=0.028, r=-0.752$) and \textit{\textbf{D}} ($W=31, p=0.041, r=-0.674$) led to significantly lower ratings than the baseline \textit{\textbf{A}} in AR presence (see Figure~\ref{fig:shape_presence}).
Regarding TAR presence, a significant influence of the shape condition could be detected neither for part 1 nor for part 2 of the study.

\subsection{Disturbances}
Regarding disturbance during grasping the objects, the Friedman test revealed significant effects both for study part 1 ($\chi^2=36.105, p<0.001$) and 2 ($\chi^2=50.119, p<0.001$). The post-hoc test then showed that conditions \textit{\textbf{E}} ($W=0, p=0.003, r=1.0$) and \textit{\textbf{D}} ($W=5, p=0.009, r=0.905$) had significantly higher values of disturbance during grasping than the baseline \textit{\textbf{A}} in part 1. For part 2, the results are similar with only conditions \textit{\textbf{E}} ($W=0, p<0.001, r=1.0$) and \textit{\textbf{D}} ($W=8, p=0.035, r=0.824$) standing out with higher ratings.
Disturbance during interaction with the objects behaves similarly. In part 1 ($\chi^2=27.594, p<0.001$) and part 2 ($\chi^2=37.407, p<0.001$), a significant influence of shape condition on the ratings was found. Again for part 1, only conditions \textit{\textbf{E}} ($W=8, p=0.003, r=0.906$) and \textit{\textbf{D}} ($W=18, p=0.031, r=0.735$) and for part 2, likewise conditions \textit{\textbf{E}} ($W=0, p=0.001, r=1.0$) and \textit{\textbf{D}} ($W=8.5, p=0.022, r=0.838$) were rated significantly higher in disturbance during interaction than condition \textit{\textbf{A}} (see Figure~\ref{fig:shape_disturbance}).
Overall, the results show a tendency for the disturbance ratings to increase with increasing level of abstraction and that small differences between the virtual and physical object are tolerated.

\subsection{Task Completion Time}
The Friedman test indicates a significant influence of the shape condition on the measured task completion times only in study part 2 ($\chi^2=18.08, p=0.001$), not in part 1 ($\chi^2=6.8, p=0.147$). Looking further into part 2, both conditions \textit{\textbf{E}} ($W=21, p=0.003, r=0.8$) and \textit{\textbf{D}} ($W=28, p=0.011, r=0.733$) showed a significantly increased task completion time compared to baseline condition \textit{\textbf{A}}.
These results imply that at least slightly abstracted proxy objects can be used without significantly degrading performance.

\subsection{Final Questionnaire}
The study participation was finished by completing a final questionnaire where participants ranked the five shape abstractions based on three criteria: feeling of realism, easiness, and overall preference.
For each ranking, the condition with the highest score was given 5 points, the second ranked condition was given 4 points, and so on until the last condition received 1 point.
With this approach, a sum of scores was computed for each shape condition which is reported in Table~\ref{tab:ranking}.
In all three categories, there is a clear trend that the position in the ranking worsens as the level of abstraction increases.

\section{Discussion}
We hypothesized that shape differences between the virtual and physical object cannot be properly detected for very similar shapes (H1). Figure~\ref{fig:shape_summary} shows that for both parts of the study conditions \textit{\textbf{B}} and \textit{\textbf{C}} were not rated significantly differently compared to the baseline condition \textit{\textbf{A}} in our study setup, while conditions \textit{\textbf{D}} and \textit{\textbf{E}} were.
This shows that many participants did not perceive small differences between the physical and the virtual objects, supporting our hypothesis H1.

We expected that small shape differences between the physical and virtual object would not be perceived. Therefore, we also assumed that the shape of the physical object could be varied to a certain degree without significantly worsening the AR presence and the TAR presence (H2).
AR presence worsened significantly only in part 2 for the levels of abstraction \textit{\textbf{D}} and \textit{\textbf{E}}. In part 1, however, no significant difference in AR presence was found for abstractions \textit{\textbf{B}} to \textit{\textbf{E}} compared to the baseline condition \textit{\textbf{A}}.
In Figure~\ref{fig:shape_presence}, a slight trend can be seen in part 1 that the presence decreases with increasing degree of abstraction. In our chosen study setup, however, no significant deterioration was detectable.
Likewise, the TAR presence scores only slightly decrease with increasing abstraction level (see Figure~\ref{fig:shape_presence}). For none of the shape conditions could a significantly lower TAR presence, compared to the baseline condition \textit{\textbf{A}}, be determined.
The results suggest that it is possible to use an object simplified in shape as a tangible proxy to manipulate a virtual object without significant degradation in presence, thus supporting our hypothesis H2.

We also hypothesized that the shape of the virtual and physical object can differ within a certain range without significantly worsening usability (H3).
Figure~\ref{fig:shape_summary} shows that in our study setup the shape could be abstracted up to the condition \textit{\textbf{C}} without a significant worsening of the grasp and interaction disturbance ratings.
The results show a clear trend that disturbance increases with increasing abstraction of the proxy object and that small abstractions are possible without significantly degrading usability, which strengthens our hypothesis H3.

In addition to shape perception, presence and disturbance, we also evaluated the influence of the shape condition on the task completion time for both parts of the study separately. Again, we assumed that shape differences between the physical and virtual object are possible to a certain degree without a significant deterioration of the task completion time, but that the choice of the degree of abstraction has an influence on the performance (H4). 
Hypothesis H4 was partially supported by our results. While for study part 1 no significant influence of the shape condition on the task completion time could be found, an influence was found for part 2. Here, the completion time worsened significantly starting from the abstraction level \textit{\textbf{D}} compared to the baseline condition \textit{\textbf{A}}.
This implies that small differences in the shape compared to the virtual overlay are possible without significantly worsening the task completion time.
Study part 2, where the assembled sofa combination had to be built, was more difficult for the participants when using the highly abstracted shapes \textit{\textbf{D}} and \textit{\textbf{E}}, which is reflected in the time taken to solve the task.
One assumption for this is that with the physical props whose shapes still made it possible to recognize and feel their orientation, it was possible to assemble the sofa combination without having to concentrate intensively on the virtual overlays. 
This is particularly the case because a sofa is an everyday object, where it is clear in which orientation two parts are logically connected to each other.

Overall, the results suggest that it is possible to use objects with abstracted shapes as physical representations of different virtual objects. However, a too-strong abstraction, which does not allow users to recognize the original shape anymore, leads to significant deteriorations regarding the perceived presence, usability and -- depending on the task -- even performance.

\begin{table}
  \centering
  \caption{Summed scores for each shape condition regarding realism, easiness and preference obtained by the rankings of all participants.}
  \label{tab:ranking}
  \begin{tabular}{l c c c c c}
    \toprule
    & ~~\textit{\textbf{A}}~~ & ~~\textit{\textbf{B}}~~ & ~~\textit{\textbf{C}}~~ & ~~\textit{\textbf{D}}~~ & ~~\textit{\textbf{E}}\\
    \midrule
    Realism~~~~ & 91 & 80 & 61 & 39 & 29 \\
    Easiness~~~~ & 83 & 73 & 61 & 53 & 30 \\
    Preference~~~~ & 82 & 71 & 59 & 54 & 34 \\
  \bottomrule
\end{tabular}
\end{table}

\section{Limitations}
We investigated the extent to which it is possible to use an object with a different shape as a tangible for the manipulation of a virtual object.
The results indicate that at least small abstractions of the shape are possible without significantly degrading performance, presence, and usability.
Currently, the results are only based on investigations with sofa elements, which were abstracted to a cuboid and further to a plate. These results have to be consolidated by further investigations with other different objects, which are abstracted to other basic shapes, e.g. a sphere or a cylinder.

In addition, it is not possible to determine from the results exactly to what degree an abstraction is possible. This can only be determined with the help of special procedures, such as the up/down staircase procedure~\cite{garcia1998forced}. However, as in this study, the results would depend on the self-defined levels of abstraction.

\section{Conclusion}
In this study, we investigated the extent to which a physical object used as a tangible for interaction with a virtual object can deviate in shape from its virtual counterpart. For this purpose we determined the influence of five different shape variations on presence, usability and performance.
The baseline condition represented a detailed replica in the form of a 3D print of the virtual object, which was abstracted in defined stages up to the base area of the virtual object.

The results imply that it is possible to use a simplified proxy object for interaction without significantly degrading performance, presence, and usability. In our study setup, this was the case up to the condition ``abstraction of a standard sofa'', where the shape was still at least roughly similar to that of the virtual object. Especially for the assembly task, the shape conditions ``cuboid'' and ``plane'' performed significantly worse than the baseline condition.

In our study, we solely examined the influence of the shape variation by keeping the size factor the same for all conditions. An exception is the height in the condition ``plane'', since this is a flat object.
If a set of acceptable physical objects is to be chosen to represent a multitude of virtual objects, then also the interplay between size and shape of the proxy object will play a crucial role, which needs to be examined in detail.

\bibliographystyle{abbrv-doi}

\bibliography{references}

\end{document}